\documentclass[11pt]{article}

\usepackage{fullpage}
\usepackage{times}
\usepackage{soul}
\usepackage{url}
\usepackage[utf8]{inputenc}
\usepackage[small]{caption}
\usepackage{graphicx}
\usepackage{amsmath}
\usepackage{booktabs}

\usepackage[ruled,vlined]{algorithm2e}
\SetArgSty{textrm}

\usepackage{enumitem}

\usepackage{libertine}

\usepackage{amssymb,amsmath,amsfonts,amstext,amsthm}

\usepackage[breaklinks=true]{hyperref}
\usepackage[svgnames]{xcolor}
\usepackage[capitalise,nameinlink]{cleveref}
\hypersetup{colorlinks={true},linkcolor={DarkBlue},citecolor=[named]{DarkGreen}}

\usepackage{tikz}  
\usetikzlibrary{arrows}
\usetikzlibrary{patterns,snakes}
\usetikzlibrary{decorations.shapes}
\tikzstyle{overbrace text style}=[font=\tiny, above, pos=.5, yshift=5pt]
\tikzstyle{overbrace style}=[decorate,decoration={brace,raise=5pt,amplitude=3pt}]
\usetikzlibrary{shapes.geometric}
\usetikzlibrary{shapes,positioning}
\usetikzlibrary{calc,positioning}
\usetikzlibrary{shapes.multipart}

\definecolor{cadmiumgreen}{rgb}{0.0, 0.42, 0.24}

\usepackage{bbm}

\theoremstyle{definition}
\newtheorem{open}{Open Problem}
\newtheorem{defn}{Definition}
\newtheorem{ex}{Example}
\newtheorem*{Algorithm}{Algorithm}

\usepackage{natbib}

\usepackage{mathtools}
\usepackage{xcolor} 

\usepackage{authblk}

\usepackage{subcaption}

\usepackage{mdframed}
\mdfsetup{linewidth=1pt}

\usepackage{xspace}
\usepackage{fancyhdr}
\usepackage{tcolorbox}

\begin{document}

\allowdisplaybreaks

\title{\bf Fair Division of Indivisible Goods: A  Survey\thanks{This work is partially supported by the ERC Advanced 
 Grant 788893 AMDROMA ``Algorithmic and Mechanism Design Research in 
 Online Markets'', the MIUR PRIN project ALGADIMAR ``Algorithms, Games, 
 and Digital Markets'', and the NWO Veni project No.~VI.Veni.192.153.}}

\author[1]{Georgios Amanatidis}
\author[2]{Georgios Birmpas}
\author[3]{Aris Filos-Ratsikas}
\author[4]{Alexandros A. Voudouris}

\affil[1]{Department of Mathematical Sciences, University of Essex, UK}
\affil[2]{Department of Computer, Control and Management Engineering, Sapienza University of Rome, Italy}
\affil[3]{Department of Computer Science, University of Liverpool, UK}
\affil[4]{School of Computer Science and Electronic Engineering, University of Essex, UK}

\date{}

\maketitle

\begin{abstract}
Allocating resources to individuals in a \emph{fair} manner has been a topic of interest since the ancient times, with most of the early rigorous mathematical work on the problem focusing on infinitely divisible resources. Recently, there has been a surge of papers studying computational questions regarding various different notions of fairness for the \emph{indivisible} case, like maximin share fairness (MMS) and envy-freeness up to any good (EFX). We survey the most important results in the discrete fair division literature, focusing on the case of additive valuation functions and paying particular attention to the progress made in the last 10 years. 
\end{abstract}

\section{Introduction}
Fair division is concerned with the fundamental task of \emph{fairly} partitioning or allocating a set of resources to a set of people with diverse and heterogeneous preferences over these resources. The associated theory originated in the works of \citet{Steinhaus49}, Banach, and Knaster (see \citep{dubins1961cut}), and has been in the focus of economics, ma\-the\-ma\-tics and computer science for the better part of the last century. Most of the classic work on the problem has been devoted to the fair division of \emph{infinitely divisible} resources, where  ``fair'' here may have different interpretations, with two predominant ones being \emph{proportionality} \mbox{\citep{Steinhaus49}} and \emph{envy-freeness} \citep{GS58,Varian74}.

Compared to the divisible setting, the fair division of \emph{indivisible} resources, referred to as \emph{discrete fair division}, turns out to be inherently more challenging. Indeed, it is clear that no reasonable fair solution can be guaranteed in some cases, e.g., when there is a single valuable item. A typical remedy to this situation is to employ ran\-do\-mi\-zation, and aim for fairness (e.g., envy-freeness) in expectation. 
A fundamentally different approach to discrete fair division came via the introduction of appropriate \emph{relaxations} of envy-freeness and proportionality, originating in the works of  \citet{LMMS04,Budish11,CaragiannisKMPS19,GourvesMT14}, which are geared to escape adverse examples. The main notions that were introduced in this literature were \emph{envy-freeness up to one good} (EF1), \emph{envy-freeness up to any good} (EFX) and \emph{maximin share fairness} (MMS). Since then, work on the topic has flourished, centered around fundamental questions about the existence and the efficient computation of allocations satisfying these or other related fairness criteria. 

More generally, over the past decade, discrete fair division had been in the epicenter of computational fair division, for several different fairness notions and a variety of different settings. In this survey, we highlight the main contributions of this literature, the most significant variants of the main setting, as well as some of the major open problems in the area.

\subsection{The setting}
For the general discrete fair division problem we consider here, there is a set $N$ of $n$ {\em agents} and a set $M$ of $m$ {\em goods} which cannot be divided or shared. Each agent $i$ is equipped with a {\em valuation function} $v_i:2^M \rightarrow \mathbb{R}_{\geq 0}$, which assigns a non-negative real number to each possible subset of items and is \emph{normalized}, i.e., $v_i(\varnothing) = 0$. In this survey we focus on the case where the valuation function of each agent $i$ is also assumed to be {\em additive}, so that $v_i(S) = \sum_{g \in S} v_i(g)$ for any subset of items $S \subseteq M$; $v_i(g)$ is used as a shortcut for $v_i(\{g\})$. Other types of valuation functions have also been studied and are briefly discussed but, unless otherwise specified, in what follows we refer to the additive case. 

An \emph{allocation} is a tuple of subsets of $M$, $A=(A_1,\ldots,A_n)$, such that each agent $i \in N$ receives the \emph{bundle} $A_i \subseteq M$, $A_i \cap A_j = \varnothing$ for every pair of agents $i,j \in N$, and $\bigcup_{i \in N}A_i = M$. The objective is to compute a {\em fair} allocation, i.e., an allocation that satisfies a desired fairness criterion. 
As already mentioned, since the early fair division literature 
there are two predominant fairness notions, namely {\em envy-freeness} 
and {\em proportionality}. An allocation is said to be envy-free if no agent believes that another agent was given a better bundle, i.e., envy-freeness depends on pairwise comparisons. 

\begin{defn}[Envy-freeness]
An allocation $A$ is {\em envy-free} if $v_i(A_i) \geq v_i(A_j)$ for every pair of agents $i, j \in N$.
\end{defn}

\noindent
On the other hand, an allocation is said to be proportional if each agent is guaranteed her \emph{proportional share} in terms of total value, independently of what others get. 

\begin{defn}[Proportionality]
An allocation $A$ is {\em proportional} if $v_i(A_i) \geq 
{v_i(M)}/{n}$ 
or every agent $i \in N$.
\end{defn}

Is not hard to see that in the additive case, if an allocation is envy-free, then it is also proportional, but the converse is not necessarily true. Envy-free or proportional allocations do not always exist in our setting. For example, consider the case of two agents and a single good that is positively valued by both agents. Since only one of the agents receives the good, the other agent gets zero value and, thus, she envies the agent with the item and also does not achieve her proportional share.

Despite this impossibility, one could still be interested in finding envy-free or proportional allocations \emph{when} they exist. Unfortunately, it turns out that the problem of even deciding whether an instance admits an envy-free (or proportional) allocation is NP-complete, which can be shown via a simple reduction from \textsc{Partition} \citep{LMMS04}. 
These straightforward impossibility results have led to the definition of multiple relaxations of these two notions, tailored for discrete fair division.

\section{Envy-Freeness up to One Good (EF1)}
\label{sec:EF1}
The first such relaxation of envy-freeness is {\em envy-freeness up to one good} (EF1), implicitly introduced by \citet{LMMS04}, but formally defined by \citet{Budish11}. According to EF1 it is acceptable for an agent $i$ to envy another agent $j$, as long as there exists a good in $j$'s bundle the hypothetical removal of which would eliminate $i$'s envy towards $j$.

\begin{defn}[EF1]
An allocation $A$ is {\em envy-free up to one good (EF1)} if, for every pair of agents $i,j \in N$, it holds that $v_i(A_i) \geq v_i(A_j \setminus \{g\})$ for some $g \in A_j$.
\end{defn}

\begin{ex} \label{ex:ef1}
To demonstrate the notion of EF1 (as well as EFX and MMS in Sections \ref{sec:EFX} and \ref{sec:MMS} later on), let us consider a simple example with three agents and five goods. The values of the agents for the goods are given in the following table:\vspace{-3pt}
\begin{center}
\begin{tabular}{c|ccccc}
        & $g_1$ & $g_2$ & $g_3$ & $g_4$ & $g_5$ \\\hline
$a_1$   &   15  & 3     & 2     &  2    &  6\\
$a_2$   &   7   & 5     & 5     &  5    &  7\\
$a_3$   &   20  & 3     & 3     &  3    &  3\\\hline
\end{tabular}
\end{center}
This instance does not admit any envy-free or proportional allocations.  To see this, observe that in any proportional allocation, agent $a_3$ must get at least $\{g_1\}$ or $\{g_2, g_3, g_4, g_5\}$. In the latter case at least one of $a_1$ and $a_2$ will get no goods, whereas in the former case $a_1$ must get at least three of the remaining four goods and $a_2$ must get at least two, which is not possible.
On the other hand, note that the allocation $A_1 = \{g_3, g_4\}$, $A_2 = \{g_2, g_5\}$,  $A_3 = \{g_1\}$ is EF1: $a_2$ and $a_3$ are not envious, and the envy of $a_1$ towards $a_2$ and $a_3$ can be eliminated by the hypothetical removals of $g_5$ from $A_2$ and $g_1$ from $A_3$ respectively. 
\hfill $\qed$
\end{ex}

There are simple, polynomial-time algorithms for computing EF1 allocations. The first such algorithm is known as {\em Envy-Cycle Elimination}, and was developed by \citet{LMMS04} several years before EF1 was formally defined.
\begin{Algorithm}[Envy-Cycle elimination]
Envy-Cycle Elimination operates in phases. In each phase, it first allocates one of the available goods to some agent that no other agent envies. Then, it looks for cycles in the current {\em envy-graph} (a graph that contains a node for each agent and a directed edge from agent $i$ to agent $j$ if and only if $i$ envies $j$), and eliminates them by appropriately reallocating the bundles of the involved agents. This guarantees that there is always an agent no one envies at the beginning of the next phase. 
\end{Algorithm}

\noindent 
While Envy-Cycle Elimination works for any monotone valuations, for the additive case EF1 allocations can be computed using a much simpler draft algorithm, known as {\em Round-Robin} \citep{CaragiannisKMPS19}. 

\begin{Algorithm}[Round-Robin]
Round-Robin fixes an ordering of the agents and, according to this ordering, it lets one agent at a time choose their favorite available good until all goods have been allocated. 
\end{Algorithm}

\noindent 
To see why Round-Robin achieves EF1 allocations, consider two agents $i$ and $j$, such that $i$ comes before $j$ in the ordering. As $i$ has the chance to pick a good before $j$ in every single round of the algorithm, $i$ cannot envy $j$. Of course, agent $j$ may envy agent $i$. Let $g$ be the first good chosen by $i$. From that point on, we can see the execution of the algorithm on the remaining goods as a fresh run where now $j$ has the chance to pick a good before $i$ in every round. So $j$ does not envy $i$'s bundle after the removal of good $g$ from it.

While EF1 allocations are rather easy to achieve, as demonstrated above, \citet{CaragiannisKMPS19} identified an interesting inherent connection between EF1 and the notion of {\em maximum Nash welfare (MNW)}.

\begin{defn}[MNW allocation]
An allocation $A$ is said to be a maximum Nash welfare (MNW) allocation if (a) it maximizes the product of agent values $\prod_i v_i(A_i)$, and (b) in case the Nash welfare of all allocations is $0$, it then maximizes the product for the agents with positive value. 
\end{defn}

\noindent 
In particular, \citet{CaragiannisKMPS19} showed that MNW allocations are always EF1 and also Pareto optimal (PO). This result shows that there exist allocations that combine fairness with other desired properties, in particular with PO. However, it is known that MNW allocations are generally hard to compute in polynomial time (in fact, it is hard to even approximate them~\citep{hoefer2021apx-mnw}). Hence, the question of whether it is possible to efficiently compute EF1 and PO allocations was left open. \citet{BarmanKV18} recently made progress by computing such allocations in pseudo-polynomial time. This brings us to our first open problem.

\begin{open}
Can an EF1 and PO allocation be computed in polynomial time?
\end{open}

\section{Envy-Freeness up to Any Good (EFX)}
\label{sec:EFX}

While EF1 is easy to achieve, in many cases it is a fairly weak fairness notion; an EF1 allocation is considered to be fair for an agent even when a very highly-valued good is hypothetically removed from another agent's bundle (e.g., a house or an expensive car). For example, consider agent $a_1$'s perspective of the allocation in Example \ref{ex:ef1}, where the proposed EF1 solution requires the removal of rather valuable goods for the agent. A very natural refinement of the notion is the stricter relaxation of {\em envy-freeness up to any good (EFX)} that was introduced in 2016 by \citet{CaragiannisKMPS19} in the conference version of their work but also somewhat earlier by \citet{GourvesMT14} under the name {\em near envy-freeness}. An allocation is said to be EFX if the envy of an agent $i$ towards another agent $j$ can be eliminated by the hypothetical removal of {\em any} good in $j$'s bundle. 

\begin{defn}[EFX]
An allocation $A$ is {\em envy-free up to any good (EFX)} if, for every pair of agents $i,j \in N$, it holds that $v_i(A_i) \geq v_i(A_j \setminus \{g\}$) for any $g\in A_j$ such that $v_i(g)>0$.
\end{defn}

\begin{ex} \label{ex:efx}
Consider again the instance of Example~\ref{ex:ef1}. The allocation $A_1 = \{g_3, g_4\}$, $A_2 = \{g_2, g_5\}$,  $A_3 = \{g_1\}$ is not EFX, since the envy of $a_1$ towards $a_2$ cannot be eliminated by removing $g_2$ ($a_1$'s least favorite good in $A_2$) from $A_2$. Nevertheless, it is easy to modify this allocation to get $B_1 = \{g_4, g_5\}$, $B_2 = \{g_2, g_3\}$,  $B_3 = \{g_1\}$ which is EFX. Indeed, the envy of $a_1$ towards $a_3$ can be eliminated by removing $g_1$ from $B_3$, whereas the envy of $a_2$ towards $a_1$ can be eliminated by removing $g_4$ from $B_1$; in both cases the hypothetical removal involves the envious agent's least valued good in the other agent's bundle. 
\hfill $\qed$
\end{ex}

In contrast to EF1, where the existence is guaranteed via simple polynomial-time algorithms, the existence of EFX allocations is a challenging open problem. \cite{procaccia2020technical} in fact referred to this as ``fair division's most enigmatic question''. In the past few years, a sequence of works have positively answered this question for important special cases, centered mainly around two axes: a small number of agents or restricted agents' valuations.

\medskip

\noindent {\em EFX for two and three agents:} \citet{PR18} showed that an EFX allocation always exists and can be efficiently computed when there are only two agents.
In a breakthrough paper, \citet{CKMS20} showed that EFX allocations always exist for instances with three agents and described a procedure that computes such an allocation in pseudo-polynomial time; computing EFX allocations for three agents in polynomial time is still an open problem.

\medskip

\noindent {\em EFX for restricted valuations:} In \citep{PR18}, it was also shown that EFX allocations exist and can be computed in polynomial time when all agents agree on the ordering of the goods with respect to their values. For instances where there are only two distinct possible values that an agent may have for each good, \citet{ABFHV21} showed that EFX allocations exist and can be efficiently computed for any number of agents; later, \citet{GargM21} showed that this is possible even in conjunction with PO. In fact, \citet{ABFHV21} also demonstrated an interesting connection between EFX and MNW allocations for bi-valued instances, by showing that MNW implies EFX. A similar result was later shown by \cite{BabaioffEF21} for general valuations with binary (i.e., $0$ or $1$) 
marginals. 

\medskip

Here, it is instructive to mention that in the related literature, the requirement that the inequality must hold only for positively-valued goods in the definition of \citet{CaragiannisKMPS19} stated above is often dropped. This stronger version of EFX is usually called EFX$_0$ \citep{kyropoulou2020groups}.
In the case of binary valuations, a special case of bi-valued instances, the distinction makes a difference but in more general settings, the existence and computation of EFX$_0$ allocations can be reduced to the existence and computation of EFX allocations; see \citep{ABFHV21} for a related discussion. It is easy to see that envy-freeness implies EFX$_0$, which implies EFX, which in turn implies EF1. 

While the aformentioned results are positive first steps towards showing the existence of EFX allocations, a general positive (or negative) result still remains elusive. This brings us to our second open problem, which is one of the most important open problems in fair division.

\begin{open}\label{open:EFX}
Does an EFX allocation exist for instances with $n \geq 4$ agents and unrestricted additive valuations?
\end{open} 

\subsection*{Relaxations of EFX}
Instead of focusing on exact EFX allocations, a growing line of work has taken a different approach by aiming to compute allocations that are \emph{approximately} EFX, for different notions of approximation. The first such notion is in terms of multiplicative approximations to the values obtained by the agents.

\begin{defn}[$\alpha$-EFX]
Let $\alpha \in (0,1]$. An allocation $A$ is $\alpha$-EFX if, for every pair of agent $i,j \in N$, it holds that $v_i(A_i) \geq \alpha \cdot v_i(A_j \setminus \{g\}$) for any $g\in A_j$ such that $v_i(g)>0$. 
\end{defn}

\noindent 
\citet{PR18} were the first to pursue this, showing that $1/2$-EFX allocations always exist, even for subadditive valuation functions, and later \citet{ChanCLW19} showed that computing such allocations can be done in polynomial time. The approximation ratio for the additive case was further improved by \citet{ANM2019} to $\phi-1 \approx 0.618$ by combining Round-Robin and Envy-Cycle Elimination with some appropriate pre-processing. To this end, we have the following open question: 

\begin{open}\label{open:alpha-EFX}
What is the best possible $\alpha$ for which $\alpha$-EFX allocations exist? 
\end{open}

A positive answer to Open Problem~\ref{open:EFX} would establish that $\alpha=1$ in Open Problem~\ref{open:alpha-EFX}, but a negative answer would make the latter open problem very meaningful in its own right. Additionally, as  is the case for all of these notions, the next natural question is whether existence can be paired with polynomial-time algorithms for finding such allocations, or whether some kind of computational hardness can be proven.

Another recent approach is that of relaxing the requirement to allocate \emph{all} available goods. Clearly, if done without any constraints, this makes the problem trivial: simply leaving all goods unallocated, results in an envy-free allocation. However, the objective here is to only leave ``a few'' goods unallocated (e.g., donate them to charity instead), or remove some goods without affecting the maximum possible Nash social welfare by ``too much''. On this front, \citet{caragiannis2019charity} showed that it is possible to compute an EFX allocation of a subset of the goods, the Nash welfare of which is at least half of the maximum Nash welfare on the original set. \citet{CKMS20} presented an algorithm that computes a partial EFX allocation, but the number of unallocated goods is at most $n-1$, and no agent prefers the set of these goods to her own bundle. 
The latter result was recently improved by \citet{BCFF21} who showed that the unallocated goods can be decreased to $n-2$ in general, and to just one for the case of $4$ agents. Finally, \citet{ChaudhuryGMMM21} showed that a $(1-\varepsilon)$-EFX allocation with at most a sublinear number of unallocated goods and high Nash welfare can be computed in polynomial time. This motivates the next open problem.

\begin{open}
Is it possible to achieve an exact EFX allocation by donating a sublinear number of goods?
\end{open} 

\section{Maximin Share Fairness (MMS)}
\label{sec:MMS}

Besides the two additive relaxations of envy-freeness discussed so far, an extensively studied fairness notion in discrete fair division is {\em maximin share fairness}, also introduced by \cite{Budish11}. The notion can be seen as a generalization of the rationale of the well-known cut-and-choose protocol, which is known to guarantee an envy-free partition of a divisible resource. 
Here the goal is to give to each agent $i$ goods of value at least as much as her \emph{maximin share} $\mu^n_i(M)$, which is the maximum value this agent could guarantee for herself by partitioning the goods into $n$ disjoint bundles and keeping the worst of them. As such, it is a relaxation of proportionality. 

\begin{defn}[MMS]\label{def:mms}
Let $\mathcal{A}_n(M)$ be the collection of possible allocations of the goods in $M$ to $n$ agents. An allocation $A$ is said to be {\em maximin share fair (MMS)} if for each agent $i \in N$, it holds that 
$v_i(A_i) \geq \mu^n_i(M) = \!\!\displaystyle\max_{B \in \mathcal{A}_n(M)} \min_{S \in B} v_i(S).$
\end{defn}

\begin{ex} \label{ex:mms}
Returning to the instance of Example~\ref{ex:ef1}, we can see that $\mu^3_1(M) = 6$, since it is not possible to partition the items into three sets with strictly more value, but $6$ is guaranteed by the partition $\{g_1\}$, $\{g_2, g_3, g_4\}$, $\{g_5\}$. Similarly, $\mu^3_2(M) = 7$ and $\mu^3_3(M) = 6$.
Therefore, $B_1 = \{g_4, g_5\}$, $B_2 = \{g_2, g_3\}$,  $B_3 = \{g_1\}$, from Example \ref{ex:efx}, is an MMS allocation, but $A_1 = \{g_3, g_4\}$, $A_2 = \{g_2, g_5\}$,  $A_3 = \{g_1\}$, from Example \ref{ex:ef1}, is only a $2/3$-MMS allocation as agent $a_1$ gets a bundle of value $4 = 2/3 \cdot \mu^3_1(M)$ (see also below).
\hfill $\qed$
\end{ex}

While it is easy to see that computing MMS allocations or even computing the maximin share of an agent is an NP-hard problem using a reduction from {\sc Partition}, there is a PTAS for the latter task \citep{Woeginger97}. 
The first breakthrough about MMS was that allocations that guarantee it do not always exist when there are more than two agents~\citep{KurokawaPW18,KPW16}, yet it is possible to compute \emph{approximate} MMS allocations. We say that an allocation is $\alpha$-MMS, for $\alpha \in (0,1]$, if each agent $i$ is guaranteed to get value at least $\alpha\,  \mu^n_i(M)$. 
To this end, \citet{KurokawaPW18} showed how to find $2/3$-MMS allocations, albeit not in polynomial time. \citet{AMNS17} matched this guarantee in polynomial time, as did \citet{BarmanK20} with a much simpler algorithm. 
The barrier of $2/3$ was broken by \citet{ghodsi2021improvement} who designed an elaborate $(3/4-\epsilon)$-approximation algorithm. A simpler algorithm with a slightly improved approximation guarantee of $3/4 + 1/(12n)$ was proposed by \citet{GT21}. On the negative side, \citet{FST21} recently showed that it is impossible to achieve an approximation bound better than $39/40$. \citet{BarmanK20} and \citet{GhodsiHSSY22} designed algorithms for computing approximate MMS allocations for richer classes of valuations (such as submodular, XOS, and subadditive).

\begin{open}
Is it possible to improve upon the bound of $3/4 + 1/(12n)$ for additive valuations? Is there a stronger inapproximability bound than $39/40$?
\end{open}

As expected, by restricting the number of agents or the space of the valuation functions, one can get stronger results.

\medskip

\noindent {\em MMS for four or fewer agents:} 
When there are only two agents a simple cut-and-choose protocol always produces an MMS allocation. Specifically, the first agent partitions the set of goods as equally as possible (thus the worst set has value equal to her maximin share) and the second agent chooses who gets each of these sets. As suggested above, the first step is computationally hard but producing a $(1-\epsilon)$-MMS allocation in polynomial time is still possible.
Even though in the general case, the algorithm of \citet{KurokawaPW18} guaranteed a $2/3$-approximation, for three or four agents it guarantees an improved $3/4$-approximation. The approximation factor for the case of three agents was then improved to $7/8$ \citep{AMNS17} and later to $8/9$ \citep{gourves2019MMS}, whereas for the case of four agents the factor was improved to $4/5$ \citep{ghodsi2021improvement}.

\medskip

\noindent {\em MMS for restricted valuations:} It follows by Definition \ref{def:mms} that MMS allocations exist for instances where all agents have identical valuation functions. \citet{BouveretL16} showed that, unlike with EFX, the hardest instances for MMS (among all possible instances) are the ones where all agents agree on the ordering of the goods. They also suggested a simple construction of exact MMS allocations when the valuation functions are binary. 
This result can be generalized, as MMS allocations always exist and can be computed efficiently for ternary valuation functions \citep{AMNS17} and for bi-valued valuation functions \citep{Feige22}. \citet{EbadianPS21} showed that this is also the case when there are at most two values per agent (possibly not common across all agents) and for general instances where the value of each good is at least as much as the value of all lesser goods combined.

\begin{open}
Are there other classes of structured valuations for which MMS is guaranteed to exist, such as when there are only a few (but more than two) possible values?
\end{open}

\section{Further Notable Fairness Notions} \label{sec:other-notions}
\noindent {\em EFL and EFR:} The EFX notion was defined as a more realistic counterpart to EF1, however, as we discussed in Section~\ref{sec:EFX}, it is still unknown if it can always be guaranteed. This has led to the definition of notions that lie ``in-between'' EF1 and EFX. \cite{BBMN18} defined the notion of {\em envy-freeness up to one less-preferred good} (EFL) according to which an agent $i$ may envy another agent $j$ if either $A_j$ contains at most one good that $i$ values positively, or the envy of $i$ can be eliminated by the hypothetical removal of a good $g \in A_j$ such that $v_i(A_i) \geq v_i(g)$. They showed that EFL allocations always exist and can be computed using a variant of Envy-Cycle Elimination. \cite{farhadi2021efr} defined the notion of {\em envy-freeness up to a random good} (EFR) according to which the envy of an agent $i$ towards another agent $j$ can be eliminated \emph{in expectation} after the hypothetical removal of a randomly chosen good from $A_j$. They showed that a $0.73$-EFR allocation can be computed in polynomial time.

\begin{open}
Does an EFR allocation always exist?
\end{open}

\noindent {\em PMMS and GMMS:} Several variations of MMS have also been considered. \citet{CaragiannisKMPS19} defined the notion of {\em pairwise maximin share fairness} (PMMS) according to which, for every pair of agents $i$ and $j$, $i$'s value for $A_i$ must be at least as much as the maximum she could obtain by redistributing the set of goods in $A_i \cup A_j$ into two bundles and picking the worst of them. In other words, instead of requiring the maximin share guarantee to be achieved for the set of all agents, PMMS requires that it is achieved for any pair of agents. Despite the apparent similarities in the definitions of PMMS and MMS, \cite{CaragiannisKMPS19} showed that their exact versions are actually incomparable. The main open problem here is the following.
\begin{open}
Does a PMMS allocation always exist?
\end{open}
\noindent
Interestingly, showing the existence of PMMS allocations is at least as hard as showing the existence of EFX allocations (Open Problem~\ref{open:EFX}), as PMMS implies EFX. For approximate PMMS, the best known result is $0.781$ by  \citet{Kurokawa17}. 

An even stronger notion, which implies both MMS and PMMS, is that of {\em groupwise maximin share fairness} (GMMS) defined by \citet{BBMN18}, and which requires that the maximin share guarantee is simultaneously achieved for any possible subset of agents. \citet{BBMN18} showed that GMMS allocations exist for some restricted settings, such as when the agents have  binary or identical values. They also showed that any EFL allocation is $1/2$-GMMS, and thus such an allocation can be computed efficiently. The currently best known approximation of GMMS is $4/7$  \citep{ANM2019,CKMS20}. The implication relations between all the aforementioned notions has been used many times to show that particular algorithms have guarantees that hold for multiple notions at once. We refer the reader to the paper of \citet{ABM18} for a discussion of the relations between (approximate versions of) these notions.

\begin{open}\label{open:alpha-GMMS}
What is the best possible $\alpha$ for which $\alpha$-GMMS allocations exist? 
\end{open}

\noindent {\em Prop1, PropX and PropM:} A line of work has also focused on relaxations of proportionality that are similar in essence to EF1 and EFX.  \citet{conitzer2017prop1} defined the notion of {\em proportionality up to one good} (Prop1) according to which each agent could obtain her proportional share if given one extra good. An allocation that is Prop1 and PO always exists \citep{conitzer2017prop1} and can be computed in polynomial time \citep{Barman2019prop1}. 
\cite{aziz2020propx} defined PropX which demands that each agent can obtain her proportional share when given the least positively-valued good among those allocated to other agents. PropX is rather demanding and it cannot be always guaranteed, even in simple scenarios. Recently, \citet{baklanov2021propm1,baklanov2021propm2} introduced the notion of {\em proportionality up to the maximin good} (PropM) and showed that such allocations always exist and can be computed in polynomial time.

\section{Other Topics}

Here we consider other interesting directions like the relation between fairness and efficiency, or the possibility to achieve fairness when the agents are strategic. Finally, we briefly discuss further meaningful discrete fair division settings. 

\subsection{Fairness and Efficiency}
There is a significant line of work that considers the question of whether it is possible to simultaneously achieve fairness and \emph{efficiency}. A common type of efficiency is that of Pareto optimality, which, as we already discussed, can be guaranteed in conjunction with some fairness notions, like EF1 and Prop1. Another natural goal is to (approximately) maximize some objective function of the values of the agents, such as the {\em social welfare}, i.e.,  the total value of the agents for the goods they receive. To this end, \citet{bertsimas2011efficiency} and \citet{caragiannis2012efficiency} defined the {\em Price of Fairness}, a measure which, similarly to the approximation ratio for algorithms, measures the deterioration of the objective due to the fairness requirement (which may refer to any fairness notion).

The Price of Fairness for EF1 and EFX allocations (only for instances with two agents) was considered by \citet{BeiLMS21}. \cite{BarmanB020} managed to close a gap on the price of EF1 that was left open in the work of \citet{BeiLMS21}, and also showed tight bounds for other fairness notions, in particular, $1/2$-MMS and Prop1. \citet{halpern2021limited} showed tight bounds on the Price of EF1 and of approximate MMS under the constraint of having only ordinal information about the agent values, a typical assumption made in the context of \emph{distortion} in social choice~\citep{distortion-survey}.

\subsection{Fair Division with Strategic Agents}
\label{subsec:strategic}
Most of the papers mentioned so far, studied the problem from an algorithmic perspective under the assumption that the agents are non-strategic. In the \emph{strategic} version of the problem, an agent may intentionally misreport how she values the goods in order to end up with a better bundle. This introduces an additional layer of difficulty, as the goal is to produce fair allocations \emph{according to the true values} of the agents, while their declarations might be far from truth. This version of the problem has been considered mostly from a mechanism design \emph{without monetary transfers} perspective, in which the utility of an agent is defined as her (true) value for her bundle. 
    
A first direction was the design of \emph{truthful} mechanisms (i.e., mechanisms where no agent has an incentive to lie) that are also fair. \citet{CKKK09}, showed that no truthful mechanism for two agents and two goods can always output allocations of minimum envy. \citet{ABM16} revisited the problem for the case of two agents and any number of goods, and showed that no truthful mechanism can always output $\alpha$-MMS allocations, for $\alpha > 2/m$. 
A characterization of truthful mechanisms for two agents, showing that truthfulness and fairness are incompatible (in the sense that there is no truthful mechanism with bounded fairness guarantees under \emph{any} meaningful fairness notion) was provided by \cite{ABCM17}. This impossibility, however, does not apply to restricted cases. \citet{0002PP020} showed that for binary valuations, there is a polynomial-time truthful mechanism that produces EF1 and PO allocations, and \citet{BabaiEF21} showed an analogous result with respect to MMS. In fact, \citet{BabaiEF21} also showed that for the submodular analog of binary valuations, there is a truthful mechanism that always outputs EFX allocations.
    
More recently, the aforementioned impossibility results led to a different direction, where the focus was shifted to the stable states of non-truthful mechanisms. In particular, \cite{ABFLLR21} studied mechanisms that always have \emph{pure Nash equilibria}, and showed that every allocation that corresponds to an equilibrium of Round Robin is EF1 with respect to the (unknown) true values of the agents. 

\begin{open}
Are there mechanisms that always have pure Nash equilibrium allocations with stronger guarantees than EF1?
\end{open}

\subsection{Notable Variants of the Setting}
\label{sec:other_settings}

We conclude by giving a brief overview of other interesting settings in discrete fair division and their main results.

\subsection*{Arbitrary Entitlements} 
So far, it is always assumed that agents have equal entitlements over the goods. However, there are settings where the fairness of an allocation must be considered with respect to asymmetric entitlements, e.g., in many inheritance scenarios, closer relatives have higher entitlements often determined by law. In order to capture fairness in the presence of arbitrary entitlements, one may generalize existing notions to their weighted counterparts, like \emph{weighted} MMS \citep{FarhadiGHLPSSY19} and \emph{weighted} EF1 \citep{chakraborty2021weighted}, or tweek their definitions approprietly, like in the MMS-inspired \emph{$\ell$-out-of-$d$ share} of \citet{BabaioffNT21}. In a recent work, \citet{BabaioffEF21} introduce the notion of \emph{AnyPrice share} (APS) as the maximum value an agent can guarantee to herself if she has a budget equal to her entitlement and the goods are adversarially priced with prices that sum up to 1, and show how to efficiently compute an allocation where everyone gets value no less than 3/5 of her APS.

\subsection*{Group Fairness}
In the model we discussed in the main part of the survey, each agent is assumed to be unrelated to other agents. However, there are applications where it makes more sense for agents to be grouped together (e.g., each group might correspond to a family). Several models capturing scenarios along these lines have been considered in the literature.  \citet{suksompong2018groups} focused on a setting where each agent derives full value from all the goods allocated to the group she belongs to, and showed bounds on the best possible approximation of MMS. \citet{kyropoulou2020groups} considered EF1 and EFX allocations in the same setting, as well as in settings with dynamic group formation; some of these results were later improved by \citet{manurangsi2021groups} using ideas from discrepancy theory. \citet{halevi2019democratic} focused on the case of democratic fairness, where the goal is to compute allocations that are considered fair (e.g., satisfying EF1) by a high fraction of the agents in each group. A different model was studied by \cite{conitzer2019groups}, where goods given to a group are then distributed among its members, and thus the agents derive value only from the goods allocated personally to them.

\subsection*{Online Fair Division} 
Our model here is static, as all items, agents, and their valuation functions do not change over time. Online fair division considers settings where the agents or the goods arrive in an online manner. In the most common model there is a fixed set of agents, items arrive one by one, and they need to be allocated to the agents immediately and irrevocably \citep{AleksandrovAGW15,aleksandrov2020onlinesurvey}. Usually, in order to bypass strong negative results and show that it is possible for envy to vanish over time or the allocations to always remain EF1, the values of the goods are assumed to be bounded \citep{BenadeKPP18,zeng2020dynamic} or a limited number of reallocations is allowed \citep{HePPZ19}. The alternative model which considers a fixed set of resources and agents who arrive or depart over time has not been considered for indivisible resources, partially because it is very challenging to achieve positive results \citep{KashPS14}.

\subsection*{Randomness in Fair Division}
Until very recently there were barely any works on randomized algorithms for discrete fair division. This is partially due to the strong general preference for deterministic algorithms / mechanisms within the Computational Social Choice community (which is well-justified in many settings, yet such a discussion is beyond the scope of this survey), but also due to the nature of the problem itself. On one hand, randomness cannot help with achieving \emph{ex-post} fairness, i.e., fairness in each resulting allocation, for any of the deterministic notions mentioned herein. On the other hand, achieving \emph{ex-ante} envy-freeness, i.e., envy freeness with respect to the expected utilities, is trivial; just allocate all the goods to a single agent uniformly at random. \citet{freeman2020rand} proposed an algorithm that achieves ex-ante envy-freeness but is also ex-post EF1, i.e., all the possible allocations it outputs are deterministically EF1. \citet{aziz2020rand} gave a simpler algorithm with the same fairness guarantees that also satisfies a weak efficiency property. Finally, in a somewhat different direction, \citet{caragiannis2021interim} studied \emph{interim} envy-freeness, a notion which lies between ex-ante and ex-post envy-freeness.

\subsection*{Subsidies}  
As we saw in Section \ref{subsec:strategic}, even in a game-theoretic setting no monetary transfers are allowed in fair division problems. Indeed, arbitrary payments would significantly alter the flavor of these problems and often go against their motivation. A recent line of work, however, considers the question of whether it is possible to pay the agents just a small amount of money (subsidy) on top of a given allocation in order to make it envy-free (when the subsidies are also taken into consideration). Allocations for which this can be done are called \emph{envy-freeable}.
\citet{halpern2019subsidies}, who introduced the problem, showed that
the total subsidy needed in order to turn an envy-freeable allocation to envy-free is at most $(n-1)m v^*$, where $v^*$ is the maximum value any agent has for any good. This upper bound was later improved to $(n-1) v^*$ by \citep{brustle2020dollar}. More recently, \citet{caragiannis2021subsidies}  provided approximation guarantees and hardness results for computing an envy-freeable allocation that minimizes the total amount of subsidies.

\subsection*{Fair Division under Constraints}
Depending on the application, some allocations may not be feasible due to various restrictions, such as connectivity, cardinality, separation, or budget constraints. Such models have recently attracted the attention of the community. Rather than referring to specific works, we point the reader to the survey of \cite{suksompong2021constraints} which discusses this part of the literature in detail. 

\subsection*{Chores and Mixed Manna}  
Beyond discrete fair division of goods that we focus on in this survey, there is a significant line of work that considers similar questions when items can be seen as \emph{chores} (which are negatively valued by the agents), or \emph{mixed manna} (a mixture of both goods and chores). As such settings are out of the scope of our survey, we refer the reader to the works of \citet{AzizCIW22,BoYX,AzizRSW17,SunCD21} and references therein.  

\bibliographystyle{plainnat}
\bibliography{references}

\end{document}